\documentclass[twocolumn,showpacs,prl,aps,nofootinbib,showkeys,unsortedaddress,10pt]{revtex4-2}

\usepackage{amsmath}
\usepackage{mathtools}
\usepackage{amssymb}
\usepackage{slashed}
\usepackage{hyperref}
\usepackage{color}
\usepackage{comment}
\usepackage{simpler-wick}

\allowdisplaybreaks

\usepackage{tikz-feynhand}

\newcommand{\be}{\begin{equation}}
\newcommand{\ee}{\end{equation}}
\newcommand{\eq}[1]{Eq.\ (\ref{#1})}

\newcommand{\infinity}{\infty}

\newcommand{\fig}[1]{Fig.\ (\ref{#1})}

\newcommand{\tr}{\mbox{tr}}

\newcommand{\ess}{\hspace{0.1em}}

\newcommand{\msbar}{$\overline{\mathrm{MS}}$}

\definecolor{darkgreen}{rgb}{0,.7,0}
\definecolor{linkblue}{rgb}{0.,0.,0.9333}

\hypersetup{
    pdffitwindow=true,      
    pdfauthor={W. A. Horowitz},     
    pdfnewwindow=true,      
    bookmarksnumbered=true,
    colorlinks=true,       
    linkcolor=red,          
    citecolor=darkgreen,        
    filecolor=magenta,      
    urlcolor=cyan,           
    menucolor=blue,
}

\AtBeginDocument{\setlength\abovedisplayskip{5pt}}
\AtBeginDocument{\setlength\belowdisplayskip{5pt}}

\begin{document}

\title{Denominator Regularization in Quantum Field Theory}

\author{W.\ A.\ Horowitz}%
\email{wa.horowitz@uct.ac.za}
\affiliation{%
 Department of Physics, University of Cape Town, Rondebosch 7701, South Africa
}

\date{\today}

\begin{abstract}
    We propose a novel regularization scheme in quantum field theory, denominator regularization (den reg).  As simple to apply as dimensional regularization, and similarly compatible with a minimal subtraction renormalization scheme, den reg manifestly 1) maintains Lorentz invariance, 2) maintains gauge invariance, 3) maintains supersymmetry, 4) correctly predicts the axial anomaly, and 5) yields Green functions that satisfy the Callan-Symanzik equation.  Den reg also naturally enables regularization in asymmetric spacetimes, finite spacetimes, curved spacetimes, and in thermal field theory.
\end{abstract}
\maketitle

\section{Introduction}
In the usual calculation of quantities in quantum field theory, in which an expansion of a time- or path-ordered exponential is evaluated order by order, contributions that correspond to loops are often formally infinite \cite{Peskin:1995ev}.  These infinities must be tamed by a regularization procedure such that final, physical quantities are finite as the regulator is smoothly removed.  Examples of such regularization procedures include: momentum cutoff, Pauli-Villars \cite{Pauli:1949zm}, dimensional regularization \cite{tHooft:1979rtg} and dimensional reduction \cite{Siegel:1979wq}, zeta and operator regularization \cite{Hawking:1976ja,McKeon:1986rc}, and analytic regularization \cite{Bollini1964,Speer:1968,Lee:1983gj}.  These procedures all have serious practical and, sometimes, conceptual shortcomings.  We propose a novel regularization procedure, ``denominator regularization,'' similar to dimensional reduction and analytic regularization, which overcomes these shortcomings, with the added advantage that the procedure also permits the calculation of higher order corrections to quantum field theoretic quantities in asymmetric spacetimes, finite spacetimes, curved spacetimes, and thermal field theory.  

The origin of the ultraviolet (UV) divergences in quantum field theories (QFT) is integration up to infinitely large values of unconstrained momenta in loop diagrams.  One may render these divergences finite trivially by imposing an upper limit momentum cutoff.  Such a regularization procedure provides a natural way to understand the renormalization group \cite{Wilson:1973jj}.  However, a finite momentum cutoff explicitly breaks Lorentz invariance and, e.g., violates the Ward identity in quantum electrodynamics (QED) \cite{Peskin:1995ev}.  In Pauli-Villars regularization, massive fictitious particles are introduced with statistics such that their associated propagators make the loop integrands go to zero fast enough that the loop integrals converge.  This procedure naturally preserves Lorentz invariance, but requires multiple particles and is cumbersome when one wishes to preserve gauge invariance \cite{Bjorken:1965zz,Slavnov:1971aw}.  

Dimensional regularization, ``dim reg,'' is by far the most common regularization procedure in QFT, almost always used in conjunction with the modified minimal subtraction (\msbar{}) renormalization scheme \cite{tHooft:1973mfk,Weinberg:1973xwm,Peskin:1995ev}.  In dim reg, the number of spacetime dimensions in the problem is analytically continued from $d$ to $d-\epsilon$, where $d$ is usually 4.  By reducing the number of dimensions, the convergence properties of the integrand are improved.  The result is expanded in powers of $\epsilon$, where $1/\epsilon$ divergences are either cancelled naturally or absorbed in the renormalization procedure.  Crucially, one performs a replacement $\ell^\mu \ell^\nu\rightarrow (1/d)\eta_{\mu\nu}\ell^2$, where $\eta_{\mu\nu}$ is the usual Minkowski metric, in the integrands of loop momentum integrals where all other dependence on $\ell$ is through the invariant $\ell^2$.  Dim reg has a number of advantages: it is often, relatively speaking, simple to implement; gauge invariance is manifestly satisfied at all orders; and minimal subtraction, which is a straightforward and transparent renormalization scheme, is trivial to implement with dim reg.  However, by changing the number of spacetime dimensions, dim reg breaks supersymmetry (SUSY) \cite{Siegel:1979wq} and unitarity \cite{Hogervorst:2015akt}.  Worse, conceptually, the analytic continuation in the number of dimensions is not consistently applied: the continuation is only applied to the spacetime index $\mu$; however, the fields still use the 4 dimensional representations of SO(1,3)\footnote{While there are generalizations to representations of SO(1,$n$), these are very complicated and technically challenging \protect\cite{Binder:2019zqc,deMelloKoch:2020roo}, thus spoiling the simplicity of the usual dimensional regularization.}.  Because dim reg relies so heavily on the ``rotational'' symmetry of spacetime, one is limited to computing finite volume effects in only highly symmetric spacetimes, for example in the calculation \cite{Zinn-Justin:2002ecy} of the finite size corrections to critical exponents, and it's unclear how to generalize dim reg to spacetimes with curvature \cite{Hawking:1976ja}.  By analytically continuing the number of spacetime dimensions, one has difficulty defining the $\gamma^5$ Dirac matrix \cite{tHooft:1972tcz,Novotny:1994yx}.  The usual BMHV choice for $\gamma^5$ in dim reg, in which an infinite set of $\gamma^\mu$ are introduced, with $\gamma^5$ anticommuting for $\mu=1,\,2,\,3,\,4$ and commuting otherwise, explicitly breaks Lorentz invariance \cite{tHooft:1972tcz,Breitenlohner:1977hr}.  Perhaps worst of all, one simply cannot define the completely anti-symmetric Levi-Civita symbol in dim reg.  This latter catastrophe makes the determination of the axial anomaly in dim reg at best extremely awkward, and generally requires abandoning manifest Lorentz invariance \cite{tHooft:1972tcz,Novotny:1994yx,Peskin:1995ev}.  

Other regularization schemes only partially address dim reg's shortcomings, or have other shortcomings of their own.  Dimensional reduction \cite{Siegel:1979wq} modifies dim reg such that only momenta are treated in $d$ dimensions, whereas $\gamma$ matrices and gauge fields remain ordinary 4-vectors.  As a result, supersymmetry is not broken in dimensional reduction; however, there are still fundamental issues in defining and using $\gamma^5$ and, as a result, dimensional reduction either does not yield the correct axial anomaly or breaks supersymmetry and gauge invariance already at one loop order \cite{Stockinger:2005gx}.  Zeta regularization was invented to allow for an unambiguous way to renormalize in curved spacetimes \cite{Hawking:1976ja}.  Operator regularization is the extension of zeta regularization to higher loops  \cite{McKeon:1986rc}.  However, zeta and operator regularization in general violate BRS symmetry \cite{Rebhan:1988ed}.  Analytic regularization analytically continues the power of the denominator in momentum space propagators \cite{Bollini1964,Speer:1968,Lee:1983gj}.  The procedure is more complicated than dim reg because the continuation of the power introduces extra terms when combining denominators with Feynman parameters; these additional terms also lead to a violation of gauge invariance \cite{Rebhan:1988ed}.

In denominator regularization, or den reg, denominators are first combined using Feynman parameters.  Then the overall power of the single denominator is analytically continued from $n$ to $n+\epsilon$, where $\epsilon$ is taken sufficiently large to ensure that the integral converges in the UV.  All subsequent manipulations---i.e.\ interchanging Feynman parameter integrations with momentum integrations, Wick rotating, and ultimately integral evaluation---are rigorously well defined and valid mathematical operations.  Like in dim reg, a fictitious scale $\mu$ is introduced to preserve the dimensions of the amplitude.  Thus the Green functions derived using den reg will manifestly and straightforwardly satisfy Callan-Symanzik equations.  The final ingredient in den reg is the slight generalization of the analytic continuation of the integrand to include in addition to the scale $\mu^{2\epsilon}$ an overall coefficient function $f_{(n,p)}$ that smoothly goes to 1 as $\epsilon\rightarrow0$.  This function depends only on $\epsilon$; the original power of the denominator, $n$; and the superficial degree of divergence of the integral, $p$.  $f_{(n,p)}$ is \emph{uniquely} specified by minimally requiring that the Laurent expansion of the amplitude in $\epsilon$ has only a simple pole at $\epsilon=0$, including in the massless limit.  Thus the physics is fixed by the analytic properties of the amplitude or, equivalently, by the requirement that the amplitude is finite for any value of the convergence factor $\epsilon>0$.  With the given scheme, gauge invariance is maintained and the axial anomaly is correctly predicted manifestly in den reg.  At all times in the calculation the number of spacetime dimensions is fixed.  As a result, SUSY is preserved \cite{Siegel:1979wq}, there is no ambiguity in the definition of $\gamma^5$, and one need only consider fields in representations of SO(1,$i$), $i\in\mathbb N$ fixed.  Since no where does den reg rely on the symmetries of the spacetime, den reg is applicable to asymmetric spacetimes, finite spacetimes, curved spacetimes, and thermal field theory.  We will show that the manipulations and results are as simple to apply and arrive at as in dim reg.

\section{Den Reg in Scalar Theories}

\begin{figure}[!t]
  \centering
  \begin{tikzpicture}[baseline=(c.base)]
    \begin{feynhand}
        \setlength{\feynhandarrowsize}{4pt}
        \setlength{\feynhanddotsize}{0mm}
        \vertex [particle] (a) at (-1.5,1) {$p_A$};
        \vertex [particle] (b) at (-1.5,-1) {$p_B$};
        \vertex [dot] (c) at (-.5,0) {};
        \vertex [dot] (d) at (.5,0) {};
        \vertex [particle] (e) at (1.5,1) {$p_1$};
        \vertex [particle] (f) at (1.5,-1) {$p_2$};
        \propag [fer] (a) to (c);
        \propag [fer] (b) to (c);
        \propag [fer] (c) to [out = 45, in = 135, looseness = 1.75, edge label = $p+k$] (d);
        \propag [antfer] (c) to [out = 315, in = 225, looseness = 1.75, edge label' = $k$] (d);
        \propag [fer] (d) to (e);
        \propag [fer] (d) to (f);
    \end{feynhand}
  \end{tikzpicture}
  \caption{
  \label{f:NLOphi}
  The $s$ channel diagram contributing to the NLO correction to $2\rightarrow2$ scattering in $\phi^4$ theory, which we define as $(-i\lambda)^2 iV\big((p_1+p_2)^2;\mu\big)$.}
\end{figure}
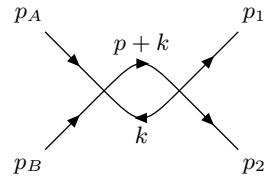

A precursor to den reg was first introduced in \cite{Horowitz:2022rpp} in the context of computing finite system size corrections to $2\rightarrow2$ scattering in massive $\phi^4$ theory at next-to-leading order (NLO).  A relevant diagram, \fig{f:NLOphi}, which is logarithmically divergent, is regulated as\footnote{There is a subleading $\mathcal O(\epsilon)$ contribution from shifting the integration variable in the logarithmically divergent integrand, which we may safely ignore.}
\newcommand{\hi}{0.5}
\newcommand{\wi}{0.25}
\newcommand{\wid}{0.55}
\begin{align}
    &V\rightarrow i\frac{({-}\mu^2)^\epsilon f_{(2,0)}(\epsilon)}{2} \int\frac{d^4k}{(2\pi)^4}\int_0^1 dx \frac{1}{(k^2-\Delta^2)^{2+\epsilon}}, \nonumber\\  
    & \Delta^2 \equiv m^2 - x(1-x)p^2-i\varepsilon, \qquad p\equiv p_1+p_2
\end{align}
The above integral is well defined for $\epsilon>0$.  We introduced, as is done in dim reg, a dimensionful scale $\mu$ that keeps the dimensions of $V$ a constant.  One should interpret the introduction of this $\mu$ as a part of the overall analytic continuation of the integrand such that the original integrand is reproduced as $\epsilon\rightarrow0$.  The minus sign multiplying $\mu$ cancels the $({-}1)^{-\epsilon}$ introduced by den reg after Wick rotation (that is absent in dim reg).  The $f_{(2,0)}$ function indicates that the original power of the denominator is 2 and the power of the numerator is such that the integral is only logarithmically divergent.  After exchanging integration orders and Wick rotating, we need the following integral:
\begin{align}
    \label{eq:logarithmic}
    \int k_E^3dk_E\frac{\mu^{2\epsilon}}{(k_E^2+\Delta^2)^{2+\epsilon}} = \frac12\frac{1}{\epsilon(1+\epsilon)}\Big(\frac{\mu^2}{\Delta^2}\Big)^\epsilon,
\end{align}
which holds for general $m^2,\,p^2\in\mathbb R$.  Similar to dim reg, the $1/\epsilon$ pole from den reg captures the logarithmic UV divergence of the integral.  

Taking $m\rightarrow0$ and integrating over the Feynman $x$ yields the beta function $B(1-\epsilon,1-\epsilon)$.  In order to minimally cancel the IR poles at $\epsilon\in\mathbb N^+$, we are uniquely led to take 
\begin{align}
    \label{eq:f20}
    f_{(2,0)}(\epsilon)\equiv\Gamma^{-2}(1-\epsilon).
\end{align}  

Expanding in powers of $\epsilon$ about $\epsilon=0$ we find
\begin{multline}
    V(p^2;\,\mu)
    \label{eq:infphi4}
    = {-}\frac{1}{2(4\pi)^2}\int_0^1dx \left[ \frac1\epsilon-1-2\gamma_E+\ln\big( \frac{\mu^2}{\Delta^2} \big) \right] \\ + \mathcal O(\epsilon), 
\end{multline}
where $\gamma_E$ is the usual Euler constant.
After slightly modifying modified minimal subtraction to subtract the constant ${-}1-2\gamma_E$, the result is identical to that from dim reg taking the usual $d=4-2\epsilon$.  Thus den reg explicitly respects unitarity to one loop in $\phi^4$ theory.  

After determining a valuable analytic continuation of the Epstein zeta function, one may compute the above NLO contribution \cite{Horowitz:2022rpp} in a system confined in a box of side lengths $L_i$, $i=1,\,2,\,3$.  After a modified minimal subtraction, the result is
\begin{multline}
    \label{e:renormfiniteV}
    \overline V(p^2,\{L_i\};\mu) = -\frac12\frac{1}{(4\pi)^2}\int_0^1dx\left\{ \ln\frac{\mu^2}{\Delta^2} \right. \\
        \left. +2\sideset{}{'}\sum_{\vec n\in\mathbb Z^3}e^{-2\pi\ess i \ess x \sum n_i p^i L_i} K_0\Big( 2\pi\sqrt{\Delta^2\sum n_i^2L_i^2} \Big)  \right\},
\end{multline}
where $K_0$ is the usual modified Bessel function.  The above can be shown to satisfy unitarity and reproduces the infinite volume result \eq{eq:infphi4} in the limit $L_i\rightarrow\infinity$ \cite{Horowitz:2022rpp}.  One may then directly compute the Callan-Symanzik equation, extract the beta function, and compute the finite system size corrections to the running coupling \cite{Horowitz:2022}.

\section{\texorpdfstring{Den Reg in Gauge Theories: \\ Boson Self Energy}{Den Reg in Gauge Theories: Boson Self Energy}}

The QED and QCD gauge boson two point functions are interesting physically because gauge invariance requires that they be transverse; i.e.\ the two point functions must satisfy the Ward identity.  Mathematically, the QED and QCD gauge boson two point functions are interesting because they involve quadratically divergent integrals.

\begin{figure}[!t]
  \centering
  \begin{tikzpicture}[baseline=(c.base)]
    \begin{feynhand}
        \setlength{\feynhandarrowsize}{4pt}
        \setlength{\feynhanddotsize}{0mm}
        \vertex [particle] (a) at (-1.5,0) {$\mu$};
        \vertex [dot] (b) at (-0.5,0) {};
        \vertex [dot] (c) at (0.5,0) {};
        \vertex [particle] (d) at (1.5,0) {$\nu$};
        \propag [pho, mom={$q$}] (a) to (b);
        \propag [pho] (c) to (d);
        \propag [fer] (b) to [out = 90, in = 90, looseness = 1.75, edge label = $q+k$] (c);
        \propag [antfer] (b) to [out = 270, in = 270, looseness = 1.75, edge label' = $k$] (c);
    \end{feynhand}
  \end{tikzpicture}
  \caption{
  \label{f:QEDNLO}
  The NLO contribution to the photon two point function in QED.  
  }
\end{figure}
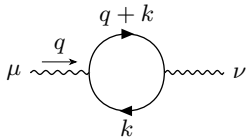

\emph{QED}. Let us now apply den reg to the photon two point function in QED to one loop, \fig{f:QEDNLO}.  One finds after combining the fermion propagator denominators, shifting the momentum integral\footnote{There is a subleading $\mathcal O(\epsilon)$ contribution from shifting the integration variable in the quadratically divergent integrand, which we may safely ignore.}, 
and discarding terms that integrate to zero by symmetry 
that for a photon of incoming momentum $q^\mu$,
\begin{align}
    &i\Pi_2^{\mu\nu} = -4e^2\int\frac{d^4\ell}{(2\pi)^4}\int_0^1dx \nonumber\\ 
    &\frac{2\ell^\mu\ell^\nu-\eta^{\mu\nu}\ell^2-2x(1-x)q^\mu q^\nu+\eta^{\mu\nu}\big(x(1-x)q^2+m^2\big)}{(\ell^2-\Delta^2)^2} \nonumber\\
    \label{eq:QED2pt}
    &\rightarrow-4e^2\int\frac{d^4\ell}{(2\pi)^4}\int_0^1dx ({-}\mu^2)^\epsilon \Big[-\frac12\eta^{\mu\nu}f_{(2,2)}(\epsilon)\ell^2 \nonumber\\
    & -f_{(2,0)}(\epsilon)[2x(1-x)q^\mu q^\nu-\eta^{\mu\nu}\big(x(1-x)q^2+m^2\big)]\Big] \nonumber\\
    & \hspace{1.75in}\times(\ell^2-\Delta^2)^{-(2+\epsilon)},\\
    &\Delta^2\equiv m^2-x(1-x)q^2-i\varepsilon. \nonumber
\end{align}
In the last line we used the symmetry of the integrand to replace $\ell^\mu\ell^\nu$ by $\eta^{\mu\nu}\ell^2/4$.  (In dim reg, the replacement is $\ell^\mu\ell^\nu\rightarrow\eta^{\mu\nu}\ell^2/d$; it is precisely this replacement that makes dim reg manifestly gauge invariant.)  Note that due to the quadratic divergence, we have a new analytic continuation function $f_{(2,2)}(\epsilon)$ and the right hand side of \eq{eq:QED2pt} is only well defined for $\epsilon>1$.

After exchanging integration orders and Wick rotating, we need the following integral
\begin{multline}
    \label{eq:quadratic}
    \int \ell_E^3 d\ell_E \frac{\mu^{2\epsilon}\ell_E^2}{(\ell_E^2+\Delta^2)^{2+\epsilon}} = {-}\frac{\Delta^2}{(1-\epsilon)\epsilon(1+\epsilon)}\Big(\frac{\mu^2}{\Delta^2}\Big)^\epsilon,
\end{multline}
which holds for general $m^2,\,p^2\in\mathbb R$.  
Notice how the quadratic divergence is captured by the $1/(1-\epsilon)$ pole, which diverges as $\epsilon\rightarrow1$, exactly at the value of $\epsilon$ at which the original integral fails to converge.  
In order to cancel this quadratic UV divergence we must take $f_{(2,2)}(\epsilon)\sim1-\epsilon$.\footnote{Note that this UV pole cancellation is equivalent to what happens manifestly in dim reg.  In dim reg, the quadratically divergent integral produces a $\Gamma\big(1-\frac 2d\big)$, which diverges logarithmically at $d=2$.  But the replacement $2\ell^\mu\ell^\nu-\eta^{\mu\nu}\ell^2\rightarrow\big(1-\frac d2\big)\eta^{\mu\nu}\ell^2/d$ gives an overall prefactor such that $\big(1-\frac 2d\big)\Gamma\big(1-\frac 2d\big)=\Gamma\big(2-\frac 2d\big)$, which cancels the logarithmic divergence at $d=2$ and effectively softens the quadratic divergence at $d=4$ to only a logarithmic divergence.}  When combined with the logarithmically divergent contributions, integration over the Feynman $x$ in the $m\rightarrow0$ limit again yields $B(1-\epsilon,1-\epsilon)$, which uniquely fixes 
\begin{align}
    \label{eq:f22}
    f_{(2,2)}(\epsilon) \equiv (1-\epsilon)\Gamma^{-2}(1-\epsilon).
\end{align}

For completeness, one finds in den reg that
\begin{align}
    \Pi_2^{\mu\nu} = -\frac{2\alpha}{\pi}\frac{\eta^{\mu\nu}q^2-q^\mu q^\nu}{\epsilon(1+\epsilon)\Gamma^2(1-\epsilon)}\int_0^1dx\,x(1-x)\Big(\frac{\mu^2}{\Delta^2}\Big)^\epsilon.
\end{align}

Note crucially the relative factor of ${-}1/2$ between the logarithmically divergent integral \eq{eq:logarithmic} and the quadratically divergent integral \eq{eq:quadratic}.  This relative factor is identical to the dim reg case.  As a result, all orders proofs of gauge invariant regularization in the den reg case will go through identically to the dim reg case.

\begin{figure}[!t]
  \centering
  \begin{tikzpicture}[baseline=(c.base)]
    \begin{feynhand}
        \setlength{\feynhandarrowsize}{4pt}
        \setlength{\feynhanddotsize}{0mm}
        \vertex [particle] (a) at (-1.5,0) {$a,\,\mu$};
        \vertex [dot] (b) at (-0.5,0) {};
        \vertex [dot] (c) at (0.5,0) {};
        \vertex [particle] (d) at (1.5,0) {$b,\,\nu$};
        \vertex [dot] (e) at (-0.433,0.25);
        \vertex [dot] (f) at (0.433,0.25);
        \propag [pho, mom={$q$}] (a) to (b);
        \propag [pho] (c) to (d);
        \propag [pho, momentum={$q+P$}] (b) to [out = 90, in = 90, looseness = 1.75] (c);
        \propag [pho, mom={$P$}] (c) to [out = 270, in = 270, looseness = 1.75] (b);
    \end{feynhand}
  \end{tikzpicture}
  \begin{tikzpicture}[baseline=(a.base)]
    \begin{feynhand}
        \vertex (a) at (-.3,0) {};
        \vertex (b) at (.3,0) {};
        \vertex (c) at (0,-.3) {};
        \vertex (d) at (0,.3) {};
        \draw (a) to (b);
        \draw (c) to (d);
    \end{feynhand}
  \end{tikzpicture}
\begin{tikzpicture}[baseline=(e.base)]
  \begin{feynhand}
    \setlength{\feynhandarrowsize}{4pt}
    \setlength{\feynhanddotsize}{0mm}
    \vertex [particle] (a) at (-1,0) {};
    \vertex [dot] (b) at (0,0) {};
    \vertex [dot] (c) at (0,1) {};
    \vertex [particle] (d) at (1,0) {};
    \vertex [dot] (e) at (0,0.25) {};
    \vertex [dot] (f) at (-0.5,0.85) {};
    \vertex [dot] (g) at (0.5,0.85) {};
    \propag [pho] (a) to (b);
    \propag [pho] (b) to (d);
    \propag [pho, style=white, mom = {[arrow style=black] $P$}] (f) to [out = 25, in = 155, looseness = 1.1] (g);
    \propag [pho] (b) to [out = 135, in = 180, looseness = 1.5] (c);
    \propag [pho] (b) to [out = 45, in = 0, looseness = 1.5] (c);
    \end{feynhand}
\end{tikzpicture}\\
  \begin{tikzpicture}[baseline=(a.base)]
    \begin{feynhand}
        \vertex (a) at (-.3,0) {};
        \vertex (b) at (.3,0) {};
        \vertex (c) at (0,-.3) {};
        \vertex (d) at (0,.3) {};
        \draw (a) to (b);
        \draw (c) to (d);
    \end{feynhand}
  \end{tikzpicture}
  \begin{tikzpicture}[
  baseline=(c.base)
  ]
    \begin{feynhand}
        \setlength{\feynhandarrowsize}{4pt}
        \setlength{\feynhanddotsize}{0mm}
        \vertex [particle] (a) at (-1.5,0) {};
        \vertex [dot] (b) at (-0.5,0) {};
        \vertex [dot] (c) at (0.5,0) {};
        \vertex [particle] (d) at (1.5,0) {};
        \propag [pho] (a) to (b);
        \propag [pho] (c) to (d);
        \propag [gho, mom=$q+P$] (b) to [out = 90, in = 90, looseness = 1.75] (c);
        \propag [gho, mom=$P$] (c) to [out = 270, in = 270, looseness = 1.75] (b);
    \end{feynhand}
  \end{tikzpicture}
  \caption{
  \label{f:YMNLO}
  The NLO contributions to the gluon two point function in Yang-Mills theory: the three gluon vertex, four gluon vertex, and ghost loop.  
  }
\end{figure}
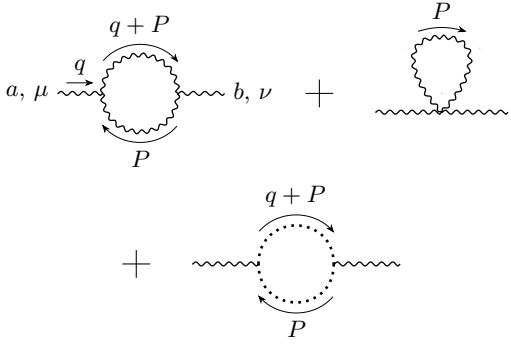

\emph{Yang-Mills}. The three diagrams contributing to the gluon two point function in Yang-Mills theory are given in \fig{f:YMNLO}.  In Feynman gauge the den reg'd three gluon vertex bubble diagram is
\begin{multline}
    -\frac{g^2}{2}C_2(G)\delta^{ab}\int\frac{d^4P}{(2\pi)^4}\int_0^1dx ({-}\mu^2)^\epsilon \Big[-\frac92f_{(2,2)}(\epsilon)\eta^{\mu\nu}P^2 \\+ f_{(2,0)}(\epsilon)\big[2q^\mu q^\nu(1+5x-5x^2)-\eta^{\mu\nu}q^2(5-2x+2x^2)\big]\Big] \\
    (P^2-\Delta^2)^{-(2+\epsilon)},
\end{multline}
the four gluon vertex bubble diagram can be manipulated to
\begin{multline}
    -g^2C_2(G)\delta^{ab}\int\frac{d^4P}{(2\pi)^4}\int_0^1dx ({-}\mu^2)^\epsilon \\ \frac{3\eta^{\mu\nu}\big( f_{(2,2)}(\epsilon)P^2 + f_{(2,0)}(\epsilon)(1-x)^2q^2 \big)}{(P^2-\Delta^2)^{2+\epsilon}},
\end{multline}
and the ghost loop diagram is
\begin{multline}
    -g^2C_2(G)\delta^{ab}\int\frac{d^4P}{(2\pi)^4}\int_0^1dx ({-}\mu^2)^\epsilon \\ \frac{\frac14\eta^{\mu\nu}f_{(2,2)}(\epsilon)P^2 - f_{(2,0)}(\epsilon)x(1-x)q^2}{(P^2-\Delta^2)^{2+\epsilon}},
\end{multline}
where $\Delta^2\equiv-x(1-x)q^2-i\varepsilon$.  We have already computed the relevant integrals, \eq{eq:logarithmic} and \eq{eq:quadratic}.  If we use the same minimal $f_{(2,0)}$ and $f_{(2,2)}$, then the full one loop Yang-Mills self energy is
\begin{multline}
    \label{eq:ymprop}
    \Pi_2^{\mu\nu \, ab} = C_2(G)\delta^{ab} \frac{\alpha_s}{4\pi}\frac{\eta^{\mu\nu}q^2-q^\mu q^\nu}{\epsilon(1+\epsilon)\Gamma^2(1-\epsilon)} \\ \times\int_0^1dx(3-4x^2)\Big(\frac{\mu^2}{\Delta^2}\Big)^\epsilon,
\end{multline}
where, by seeing that $\Delta^2$ is symmetric under $x\leftrightarrow1-x$, we have replaced $x\rightarrow1/2$ in the numerator.  Note that the $3-4x^2$ makes the integral over the Feynman $x$ in the $m\rightarrow0$ limit more complicated than in the QED case, but does not introduce any new poles in $\epsilon$.

We thus see that the exact same choices for the analytic continuation of the divergent integrals in the photon two point function, which were dictated by the analytic structure of the two point function in $\epsilon$, automatically yield a gauge invariant non-Abelian two point function (at one loop order).  

Expanding in $\epsilon$ and performing a modified minimal subtraction, one sees that the NLO Yang-Mills self-energy in den reg is identical to that given by dim reg.

\section{Axial Anomaly}
Physically, the axial anomaly is interesting because no previous regularization scheme naturally, manifestly yields the correct value for the anomaly.  (As an example, in dim reg, the definition of $\gamma^5$ in $d\ne4$ dimensions is ambiguous.  The usual choice made breaks manifest Lorentz invariance \cite{Peskin:1995ev,tHooft:1972tcz}.  In any case, computing the axial anomaly in dim reg is at best extremely subtle \cite{Novotny:1994yx}.)  Mathematically, the anomaly is interesting because the amplitude involves (superficially) linear divergences.  

\begin{figure}[!t]
  \centering
  \begin{tikzpicture}[baseline=(a.base)]
    \begin{feynhand}
        \setlength{\feynhandarrowsize}{4pt}
        \setlength{\feynhanddotsize}{0mm}
        \vertex [particle] (a) at (-2,0) {$q_\mu$};
        \vertex [dot] (b) at (-1,0) {};
        \vertex [dot] (c) at (0,0.577) {};
        \vertex [dot] (d) at (0,-0.577) {};
        \vertex [particle] (e) at (1,0.577) {$p_\nu$};
        \vertex [particle] (f) at (1,-0.577) {$k_\lambda$};
        \propag [pho] (a) to (b); 
        \propag [fer] (b) to [edge label = $\ell+p$] (c);
        \propag [fer] (c) to [edge label = $\ell$] (d);
        \propag [fer] (d) to [edge label = $\ell-k$] (b);
        \propag [pho] (c) to (e);
        \propag [pho] (d) to (f);
    \end{feynhand}
  \end{tikzpicture}
  \begin{tikzpicture}[baseline=(a.base)]
    \begin{feynhand}
        \vertex (a) at (-.3,0) {};
        \vertex (b) at (.3,0) {};
        \vertex (c) at (0,-.3) {};
        \vertex (d) at (0,.3) {};
        \draw (a) to (b);
        \draw (c) to (d);
    \end{feynhand}
  \end{tikzpicture}
  \begin{tikzpicture}[baseline=(a.base)]
    \begin{feynhand}
        \setlength{\feynhandarrowsize}{4pt}
        \setlength{\feynhanddotsize}{0mm}
        \vertex [particle] (a) at (-2,0) {$q_\mu$};
        \vertex [dot] (b) at (-1,0) {};
        \vertex [dot] (c) at (0,0.577) {};
        \vertex [dot] (d) at (0,-0.577) {};
        \vertex [particle] (e) at (1,-0.577) {$k_\lambda$};
        \vertex [particle] (f) at (1,0.577) {$p_\nu$};
        \propag [pho] (a) to (b); 
        \propag [fer] (b) to [edge label = $\ell+k$] (c);
        \propag [fer] (d) to [edge label = $\ell-p$] (b);
        \propag [pho] (c) to (e);
        \propag [pho, top] (d) to (f);
        \propag [fer] (c) to [edge label' = $\ell$] (d);
    \end{feynhand}
  \end{tikzpicture}
  \caption{
  \label{f:triangle}
  The leading order diagrams that contribute to the axial anomaly.  The incoming photon momentum $q_\mu$ is treated as an off-shell, internal line.  The outgoing photon momenta $p_\nu$ and $k_\lambda$ are on-shell.
  }
\end{figure}
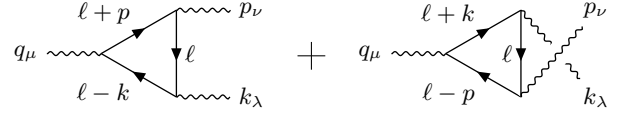

One of the two triangle diagrams with incoming, off-shell photon momentum $q^\mu$ and outgoing, on-shell photon momenta $p^\nu$ and $k^\lambda$, shown in \fig{f:triangle}, gives
\begin{align}
    &i\mathcal M^{\mu\nu\lambda}(p,k) \nonumber\\
    &= e^2 \int \frac{d^4\ell}{(2\pi)^4}\tr \Big[ \gamma^\mu \gamma^5\frac{i}{\slashed \ell - \slashed k - m + i\varepsilon} \nonumber\\
    & \qquad\qquad\qquad\qquad \gamma^\lambda\frac{i}{\slashed \ell-m+i\varepsilon}\gamma^\nu\frac{i}{\slashed \ell+\slashed p -m+i\varepsilon} \Big] \nonumber\\
    &={-}ie^2\int \frac{d^4\ell}{(2\pi)^4}\Big\{ 
    {-}\tr[\mu5\lambda\nu\alpha](\ell^2-m^2)(\ell_\alpha+p_\alpha-k_\alpha) \nonumber\\
    &\qquad\qquad+2\tr[\mu5\beta\nu\alpha]p_\alpha\ell^\lambda\ell_\beta-2\tr[\mu5\alpha\lambda\beta]k_\alpha\ell^\nu\ell_\beta \nonumber\\
    &\qquad\qquad\qquad\qquad-\tr[\mu5\alpha\lambda\beta\nu\gamma]k_\alpha\ell_\beta p_\gamma \Big\}\nonumber\\
    \label{eq:axial1}
    &\times\frac{1}{[(\ell-k)^2-m^2+i\varepsilon][\ell^2-m^2+i\varepsilon][(\ell+p)^2-m^2+i\varepsilon]},
\end{align}
where for simplicity we show only the contravariant indices of the Dirac $\gamma$ matrices in the traces.  The other triangle diagram contribution can be found from $i\mathcal M^{\mu\lambda\nu}(k,p)$, known as Bose symmetrization.  We wish to compute the axial Ward anomaly $iq_\mu i\mathcal M^{\mu\nu\lambda}$ and the vector Ward identities $p_\nu\mathcal M^{\mu\nu\lambda}$ and $k_\lambda\mathcal M^{\mu\nu\lambda}$.  One can easily show that $p_\mu,\,p_\nu,\,k_\mu,$ or $k_\lambda$ dotted into the last term in the curly bracket of \eq{eq:axial1} is zero.  Thus for our purposes we may safely ignore this term.  To evaluate the other terms, we require a number of new integrals.  After combining denominators and analytically continuing, we have for the most important integrals\footnote{We have suppressed all contributions $\mathcal O(\epsilon)$ in the following}
\begin{align}
    I_0 & \equiv -i\int \frac{d^4\ell}{(2\pi)^4}\int_0^1dx\int_0^{1-x}dy \nonumber\\
    & \qquad\qquad\frac{2f_{(3,0)}(\epsilon)(-\mu^2)^\epsilon\ell^2}{[\ell^2+2xp\cdot\ell-2yk\cdot\ell-m^2+i\varepsilon]^{3+\epsilon}}, \\
    & = \frac{2\pi^2}{(2\pi)^4}\int_0^1dx\int_0^{1-x}dy\tilde I_0, \nonumber\\
    \tilde I_0 & \equiv \frac{f_{(3,0)}(\epsilon)}{(1+\epsilon)(2+\epsilon)}\left[ \frac2\epsilon+\frac{2xyp\cdot k}{\Delta^2} \right]\left( \frac{\mu^2}{\Delta^2} \right)^\epsilon \\
    I_{0,\alpha\beta} & \equiv -i\int \frac{d^4\ell}{(2\pi)^4}\int_0^1dx\int_0^{1-x}dy \nonumber\\
    & \qquad\quad\frac{2f_{(3,0)}(\epsilon)(-\mu^2)^\epsilon\ell_\alpha\ell_\beta}{[\ell^2+2xp\cdot\ell-2yk\cdot\ell-m^2+i\varepsilon]^{3+\epsilon}}, \\
    & = \frac{2\pi^2}{(2\pi)^4}\int_0^1dx\int_0^{1-x}dy\tilde I_\eta\eta_{\alpha\beta}+\tilde I_{xy}(p_\alpha k_\beta+p_\beta l_\alpha) \nonumber\\
    & \hspace{1.25in} -\tilde I_{x^2}p_\alpha p_\beta + \tilde I_{y^2}k_\alpha k_\beta, \\
    \tilde I_{\eta} & \equiv \frac{f_{(3,0)}(\epsilon)}{(1+\epsilon)(2+\epsilon)}\left[ \frac{1}{2\epsilon}\right]\left( \frac{\mu^2}{\Delta^2} \right)^\epsilon, \\
    \tilde I_{x^iy^j} & \equiv \frac{f_{(3,0)}(\epsilon)}{(1+\epsilon)(2+\epsilon)}\left[ \frac{x^iy^j}{2\epsilon}\right]\left( \frac{\mu^2}{\Delta^2} \right)^\epsilon, \\
    \label{eq:lindiv}
    I_{1,\alpha} & \equiv -i\int \frac{d^4\ell}{(2\pi)^4}\int_0^1dx\int_0^{1-x}dy \nonumber\\
    & \qquad\qquad\frac{2f_{(3,1)}(\epsilon)(-\mu^2)^\epsilon\ell^2\ell_\alpha}{[\ell^2+2xp\cdot\ell-2yk\cdot\ell-m^2+i\varepsilon]^{3+\epsilon}}, \\
    & = \frac{2\pi^2}{(2\pi)^4}\int_0^1dx\int_0^{1-x}dy\tilde I_1(p_\alpha-k_\alpha), \nonumber\\
    \label{eq:lindivres}
    \tilde I_1 & \equiv f_{(3,1)}(\epsilon)\Big\{ \frac x2 \nonumber\\
    &\quad\quad\quad- \frac{x}{(1+\epsilon)(2+\epsilon)}\Big[ \frac3\epsilon + \frac{2xyp\cdot k}{\Delta^2} \Big]\Big( \frac{\mu^2}{\Delta^2}\Big)^\epsilon\Big\}, \\
    \Delta^2 & \equiv m^2-2xyp\cdot k -i\varepsilon.
\end{align}

Of the above integrals, the only non-trivial one is the (superficially) linearly divergent one, \eq{eq:lindiv}.  
Integrals that diverge like odd powers in the UV are subtle because shifting the dummy integration variable can leave a finite remainder \cite{Bell:1969ts,Treiman:1972}.  Following \cite{Treiman:1972}, one may derive the exact result
\begin{multline}
    \label{eq:shift}
    S^\mu(a)\equiv{-}i\int\frac{d^4\ell}{(2\pi)^4}\left( \frac{\ell^2(\ell^\mu+a^\mu)}{[\ell+a)^2-M^2+i\varepsilon]^3} \right. \\ \left. - \frac{\ell^2\ell^\mu}{[\ell^2-M^2+i\varepsilon]^3} \right) = \frac{2\pi^2}{(2\pi)^4}\frac14 a^\mu.
\end{multline}
If one evaluates $S^\mu$ in den reg, one can show that $S^\mu_{den}(a;\,\epsilon) = \frac{2\pi^2}{(2\pi)^4}\frac14 a^\mu + \mathcal O(\epsilon)$.  
As an aside, shockingly, if one computes $S^\mu$ in dim reg, one finds that the result is \emph{identically zero} for any $d<4$;
i.e.\ even though the integrands in the dimensionally regularized result smoothly go to the integrands of \eq{eq:shift} as $\epsilon\rightarrow0$, it is \emph{impossible} to correctly derive \eq{eq:shift} by smoothly taking the $\epsilon\rightarrow0$ limit of the integral, $S^\mu_{dim}$.  The origin of this discontinuity is the replacement $\ell^\mu\ell^\nu\rightarrow(1/d)\eta^{\mu\nu}\ell^2$.  For $d=4$ there is an exact cancellation that reduces the highest power in the numerator by 1.  For $d<4$ there is no highest power cancellation, and the extra contribution to the integral \emph{exactly} cancels all the other contributions.

\eq{eq:lindivres} contains only a $1/\epsilon$ pole, indicating that \eq{eq:lindiv} is actually only logarithmically divergent.  Thus none of the $f_{(n,p)}$ have contributions from UV poles and are fixed entirely by the $m\rightarrow0$ IR physics.  In order to minimally cancel the IR poles from the beta functions that result from integrating over the Feynman $x$ and $y$ parameters when $m=0$, we are uniquely led to
\begin{align}
    f_{(3,0)} & \equiv \Gamma^{-2}(1-\epsilon) \\
    f_{(3,1)} & \equiv \Gamma^{-1}(1-\epsilon)\Gamma^{-1}(2-\epsilon).
\end{align}
Then, automatically and straightforwardly, one finds that
\begin{align}
    p_\nu i\mathcal M^{\mu\nu\lambda} = k_\lambda i\mathcal M^{\mu\nu\lambda} = \mathcal O(\epsilon),\qquad m\ge 0 \\
    i q_\mu i\mathcal M^{\mu\nu\lambda} = \frac{e^2}{4\pi^2}\epsilon^{\alpha\lambda\beta\nu}k_\alpha p_\beta + \mathcal O(\epsilon),\qquad m = 0.
\end{align}
Thus den reg manifestly maintains gauge invariance for all $m$ and gives the correct $m=0$ axial anomaly \cite{Peskin:1995ev,Bell:1969ts,Treiman:1972}.

\section{Conclusions and Outlook}

We introduced the denominator regularization scheme in quantum field theory.  In den reg, the power of the denominator in an amplitude, after combined by Feynman parameters, is analytically continued; the number of spacetime dimensions is always fixed.  
The main ingredient to den reg is the further analytic continuation of the integrand by coefficient functions $f_{(n,p)}(\epsilon)$ that depend on the original power $n$ of the denominator and of the superficial degree of divergence of the integral, $p$.  The $f_{(n,p)}$ are uniquely fixed by minimally requiring that the Laurent expansion of the amplitude is free of poles other than a single $1/\epsilon$; i.e.\ we choose $f_{(n,p)}$ such that the amplitude converges for all $\epsilon>0$.  Crucially, we require the $f_{(n,p)}$ to cancel UV poles of the form $(\frac p2-\epsilon)^{-1}$ \emph{and} IR poles that emerge for $\epsilon\in\mathbb N^+$ when the theory is massless.  With this scheme in place, we showed that den reg manifestly maintains Lorentz and gauge invariance, preserves SUSY, and manifestly predicts the axial anomaly.  As in dim reg, a fictitious scale $\mu$ is introduced to keep the dimensions of amplitudes fixed.  Thus Green functions computed in den reg satisfy the Callan-Symanzik equation.  No other regularization scheme satisfies all these criteria.

The integrals that emerge in den reg are of a type and difficulty similar to dim reg.  As a result, den reg is thus well suited, as done here, to the minimal subtraction renormalization scheme, and the determination of the $f_{(n,p)}$ is straightforward and easy.  So far, the $f_{(n,p)}$ appear to be universal; once fixed for an integral in one amplitude, the same $f_{(n,p)}$ emerges from other amplitudes.  We are thus led to speculate that the $f_{(n,p)}$ are, in fact, universal.  

It will be interesting to apply den reg to other divergent amplitudes.  Den reg and dim reg will give manifestly different results for linearly and higher divergent amplitudes in scalar theories: in dim reg, these divergences will keep their same superficial degree of divergence, while den reg will soften all UV divergences to logarithmic.  We speculate that den reg will manifestly conserve the energy momentum tensor and simultaneously manifestly predict the correct Weyl anomaly.  We have checked that den reg correctly reproduces the leading order results in thermal field theory \cite{Kapusta:2006pm,Laine:2016hma}.  It will also be interesting to apply den reg to Chern-Simons theory \cite{Chaichian:1998tf}, quantities in curved spacetime \cite{Hawking:1976ja}, and to compute the finite size corrections to running couplings \cite{Horowitz:2022} and critical exponents in asymmetric systems, especially in the universality class of $\phi^4$ theory through the resummed two point function \cite{Cardy:1996xt}.  The latter may provide valuable insight, e.g., in detecting the critical endpoint of the QCD phase diagram from measurements of particle fluctuations in hadronic collisions \cite{Stephanov:1998dy,Luo:2017faz}.

\section*{Acknowledgments}
The author wishes to thank the South African National Research Foundation and the SA-CERN Collaboration for support.  The author wishes to thank Jean Du Plessis, Matthew Sievert, Alexander Rothkopf, Robert de Mello Koch, Kurt Hinterbichler, Bowen Xiao, Jonathan Shock, and Kevin Bassler for valuable discussions.

\bibliography{DRrefs}
\bibliographystyle{iopart-num}
\end{document}